\documentclass{article}

\usepackage{arxiv}

\usepackage[utf8]{inputenc} 
\usepackage[T1]{fontenc}    
\usepackage{hyperref}       
\usepackage{url}            
\usepackage{booktabs}       
\usepackage{amsfonts}       
\usepackage{nicefrac}       
\usepackage{microtype}      
\usepackage{lipsum}
\usepackage{graphicx}
\graphicspath{ {./images/} }
\usepackage{bm}
\usepackage{mathcomp}

\title{Sergeants and Soldiers in Chiral Nematic Liquid Crystal}

\author{
 Yoshiaki Uchida \\
  Graduate School of Engineering Science\\
  Osaka University\\
  1-3 Machikaneyama, Toyonaka, Osaka 560-8531, Japan\\
  \texttt{y.uchida.es@osaka-u.ac.jp} \\
   \And
 Go Watanabe \\
  Department of Data Science, School of Frontier Engineering, \\
  Kitasato University\\
  Kanagawa 252-0373, Japan\\
  \texttt{go0325@kitasato-u.ac.jp} \\
}

\begin{document}
\maketitle
\begin{abstract}
This study explores the mechanisms behind the helical structures in cholesteric liquid crystalline (CLC) phases using molecular dynamics simulations. By adding chiral agents to the nematic liquid crystalline (NLC) compound, 5CB, the research examines how the shape and motion chirality of the agents influence the overall twisting behavior. The results show that chiral agents induce shape chirality in the 5CB molecules, but dynamic chirality is not detectable at this stage. The study suggests that equilibrium is required for motion chirality to become evident, highlighting the need for further investigation.
\end{abstract}

\section{Introduction}
Helical structures appear at various levels of objects \cite{2024lago-silva_Stimuliresponsivesynthetichelicalpolymers,2024dreyfus_Dexteroushelicalmagneticrobotimprovedendovascularaccess, 2023rey-tarrio_Multichiralmaterialscomprisingmetallosupramolecularcovalenthelicalpolymerscontainingfiveaxialmotifshelix, 2023aizawa_Enantioselectivitydiscretizedhelicalsupramoleculeconsistingachiralcobaltphthalocyanineschiralinducedspinselectivityeffect}. The way in which the asymmetry that is the origin of the structure induces the overall twisting varies depending on the level. Helical structures originating from asymmetric carbons form a huge group of organic materials. Among the many organic materials, cholesteric liquid crystalline (CLC) materials show off the presence of their helical structure in the form of selective reflection \cite{2023ren_SingleEnantiomerEmitterEnabledSuperstructuralHelixInversionUpconvertingDownshiftingLuminescenceBidirectionalCircularPolarization}. The selective reflection color of cholesteric liquid crystals, which also exist in nature, is essential for the survival of living organisms. CLC materials are made up of uniaxially oriented nematic liquid crystalline (NLC) materials used in liquid crystal displays, chiral compounds that induce asymmetry, what are called, chiral agents. NLC molecules have a structure that can be approximated as a rod shape, as shown in Figure 1a. There are countless known examples of both NLC materials and chiral compounds. To understand the cause of the twisting in cholesteric liquid crystals, it is easy to intuitively understand that the asymmetrical building blocks are stacked, although this is not entirely accurate \cite{2002gennes_PhysicsLiquidCrystals}. The top and bottom faces of the rectangular prism each have grooves and protrusions of the same shape, which are slightly twisted, and when these fit together, the rectangular prism twists and aligns. This is easy to understand, but it is clear that it is inaccurate when you consider the relationship with physical properties. How can we make this easy-to-understand picture more accurate?

\begin{figure} 
  \centering
  \includegraphics[scale=0.7]{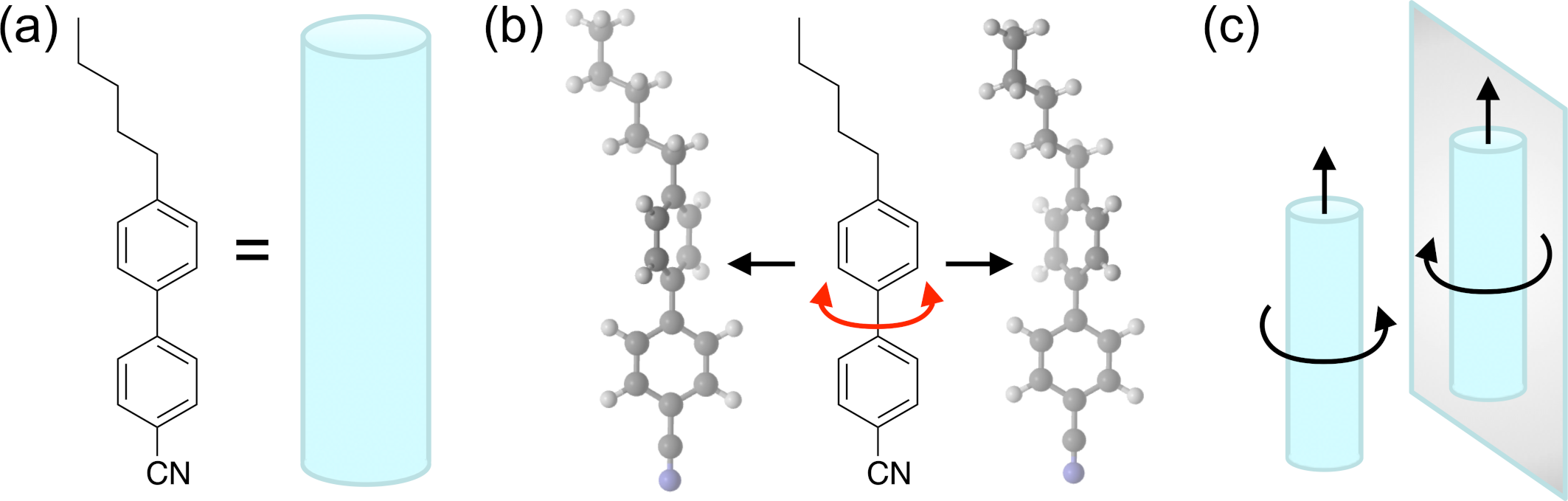}
  \caption{Two types of chirality for 5CB. (a) Cylindrical approximation of a 5CB molecule showing a nematic phase. (b) Two stable dihedral angles of the central bond of the biphenyl group of 5CB. Left and right ones are $P$ and $M$ enantiomers, respectively. (c) Chirality of dynamics for cylindrical particles. Given the combination of momentum and angular momentum, it cannot be superimposed on the mirror image.}
  \label{fig:fig1}
\end{figure}

The twisting of the molecular surface can be quantitatively expressed using the chirality parameter, which is composed of the surface tensor and helicity tensor \cite{1998ferrarini_assessmentmolecularchirality, 2016watanabe_MolecularDynamicsApproachPredictingHelicalTwistingPowersMetalComplexDopantsNematicSolvents}. This allows us to discuss the chirality of chiral agents quantitatively. However, some phenomena cannot be understood well using this alone. How is the phenomenon of the twisting of the structure of liquid crystals reversing with temperature caused? How does a CLC phase, which contains only a few percent of chiral molecules, exhibit a uniform helical structure? In the former case, the chirality of the contained molecules does not change, but the overall structure reverses. In the latter case, it is known that the molecular orientation, director, is uniformly twisted, but how do we fill the gap between this and the fact that the molecules are not asymmetric, like building blocks? After all, we have not yet grasped what the equivalent of asymmetry in building blocks is.

What is the decisive factor that determines the strength and direction of the helical structures in CLC phases? This is a classic but important question that still puzzles us. However, there are some hints. Which asymmetry of the molecules affects the overall twist? One hypothesis is that achiral substances that come into contact with chiral agents also take on a chiral shape, as shown in Figure 1b. This is known as the Sergeants and Soldiers Effect and is well known in supramolecular chemistry \cite{2010anderson_sergeantsandsoldierseffectchiralamplificationnaphthalenediimidenanotubes}. Although liquid crystals are not that rigid, the information of chirality might accumulate in the rigid core. It is necessary to confirm this. And the other is dynamical chirality \cite{2024zhou_Differentiatingenantiomersdirectionalrotationionsmassspectrometer}. Regarding the chirality of inorganic materials, which has recently attracted attention \cite{2023ishito_TrulychiralphononsaHgS, 2024oishi_Selectiveobservationenantiomericchiralphononsensuremathalphaquartz}, chirality is determined by the scalar product of momentum and angular momentum \cite{2022kishine_DefinitionChiralityEnantioselectiveFields}. Since the molecular motion in CLC phases is very disordered, it is not certain whether such a dynamic effect can overcome the noise, as shown in Figure 1c. However, the possibility that random intermolecular contact in CLC phases can create order has been confirmed in terms of magnetic properties \cite{2018nakagami_MolecularMobilityEffectMagneticInteractionsAllOrganicParamagneticLiquidCrystalNitroxideRadicalHydrogenBondingAcceptor, 2020uchida_ThermalMolecularMotionCanAmplifyIntermolecularMagneticInteractions}. This also needs to be confirmed.

Here, we will discuss the mechanism by performing MD simulations of CLC mixtures. First, we analyze the results of the MD simulations for the most well-known NLC material, 4-cyano-4'-pentylbiphenyl (5CB), shown in Figure 1, doped with a chiral agent commonly used in practice shown in Figure 2. Then, we will quantify the chirality of the molecular motion and the chirality of the molecular shapes of the constituent compounds.

\begin{figure} 
  \centering
  \includegraphics[scale=0.3]{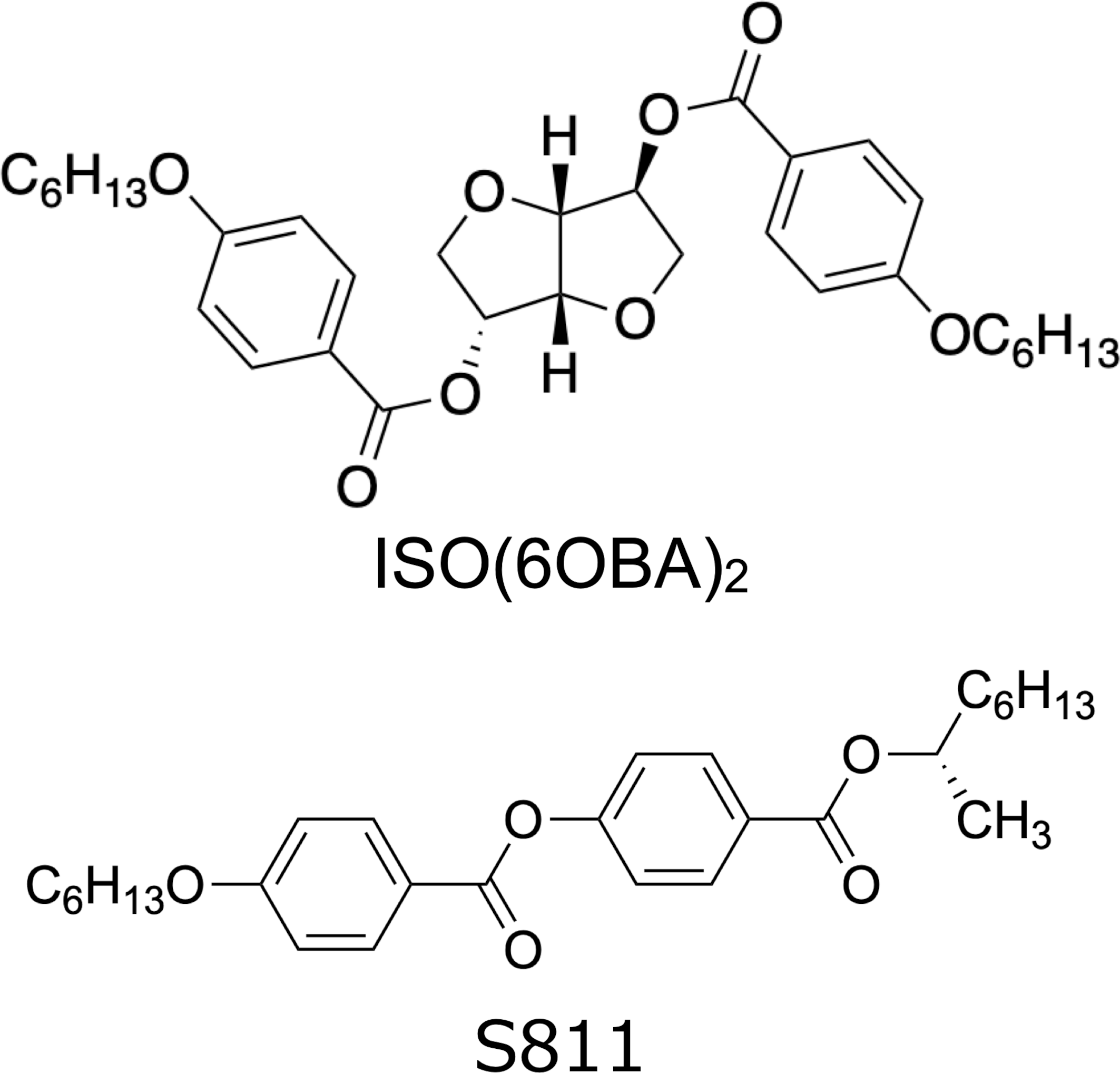}
  \caption{Molecular structures of chiral agents: ISO(6OBA)${}_2$ and S811.}
  \label{fig:fig2}
\end{figure}

\section{Methods}
\label{sec:headings}
\subsection{Molecular Dynamics Simulation}
We calculated a conformational ensemble for 5CB at the B3LYP/6-31G(d,p) level using the GAUSSIAN 16 package \cite{2019j.m.frisch_Gaussian16RevisionC02}. The dihedral angle of the central bond of the biphenyl group of 5CB was changed by $1\tcdegree$ each and fixed, and structural optimization was performed.

All-atom MD simulation was carried out using the MD programs GROMACS 4.6.6, 2016.3 and 2016.5. The partial atomic charges of the simulated LC molecule were determined by the restrained electrostatic potential (RESP)\cite{1993bayly_wellbehavedelectrostaticpotentialbasedmethodusingchargerestraintsderivingatomicchargesRESPmodel} methodology at the UB3LYP/6-31G(d,p) level using the GAUSSIAN 09 package.\cite{2009m.j.frisch_Gaussian09RevisionD01} In order to calculate the intra- and intermolecular interactions, generalized Amber force field\cite{2004wang_Developmenttestinggeneralamberforcefield} parameters were used.

The initial structure of the MD simulation system was the cubic simulation box with a side of 13.00 nm containing 2048 molecules by replicating the small rectangular cell containing both enantiomers with random rotation. 4-cyano-4’-pentylbiphenyl (5CB) was placed in the small rectangular unit cell. Chiral agents were inserted to be about 1 wt$\%$. The relaxation runs at 250 K for 5 ns, at 350 K for 1 ns and 300 K for 5 ns were successively performed. During the relaxation runs, the Berendsen thermostat and barostat \cite{1984berendsen_Moleculardynamicscouplingexternalbath} were used to keep the temperature and pressure of the system with relaxation times of 0.2 and 2.0 ps, respectively. After the relaxation runs, the equilibration run at 300 K was done for 200 ns using Nos\'e-Hoover thermostat \cite{1984nose_moleculardynamicsmethodsimulationscanonicalensemble, 1985hoover_CanonicaldynamicsEquilibriumphasespacedistributions} and Parrinello-Rahman barostat \cite{1981parrinello_Polymorphictransitionssinglecrystalsnewmoleculardynamicsmethod} with relaxation times of 1.0 and 5.0 ps, respectively. The time step was set to 2 fs since all bonds connected to hydrogen atoms were constrained with the LINCS algorithm.\cite{1997hess_LINCSlinearconstraintsolvermolecularsimulations} The smooth particle-mesh Ewald (PME) method was employed to treat the long-range electrostatic interactions, and the real space cutoff and the grid spacing are 1.4 and 0.30 nm, respectively.

\section{Results and discussion}
\label{sec:headings}
\subsection{Shape chirality of 5CB}
First, we calculate the energies of conformations with dihedral angles of the central bond of the biphenyl group of 5CB from $-90\tcdegree$ to $90\tcdegree$ using the GAUSSIAN 16 package. There are two types of energy peaks at $0\tcdegree$ and $\pm90\tcdegree$ with the most stable dihedral angles at $36\tcdegree$ and $-36\tcdegree$, as shown in Figure 2a. These peaks correspond to the $P$ and $M$ enantiomers. The energy barrier that inhibits the transition between them is also found to be $3.12 k_{\rm B}T$ at 300 K. This suggests that the chirality, which appears transiently as a conformational bias of the molecule, is well retained at room temperature. We calculated the Boltzmann distribution at 300 K using the strain energies, as shown in Figure 2b. The frequency is concentrated around two stable dihedral angles. It indicates that we can discuss the shape chirality by considering a transition between two states.

\begin{figure} 
  \centering
  \includegraphics[scale=0.7]{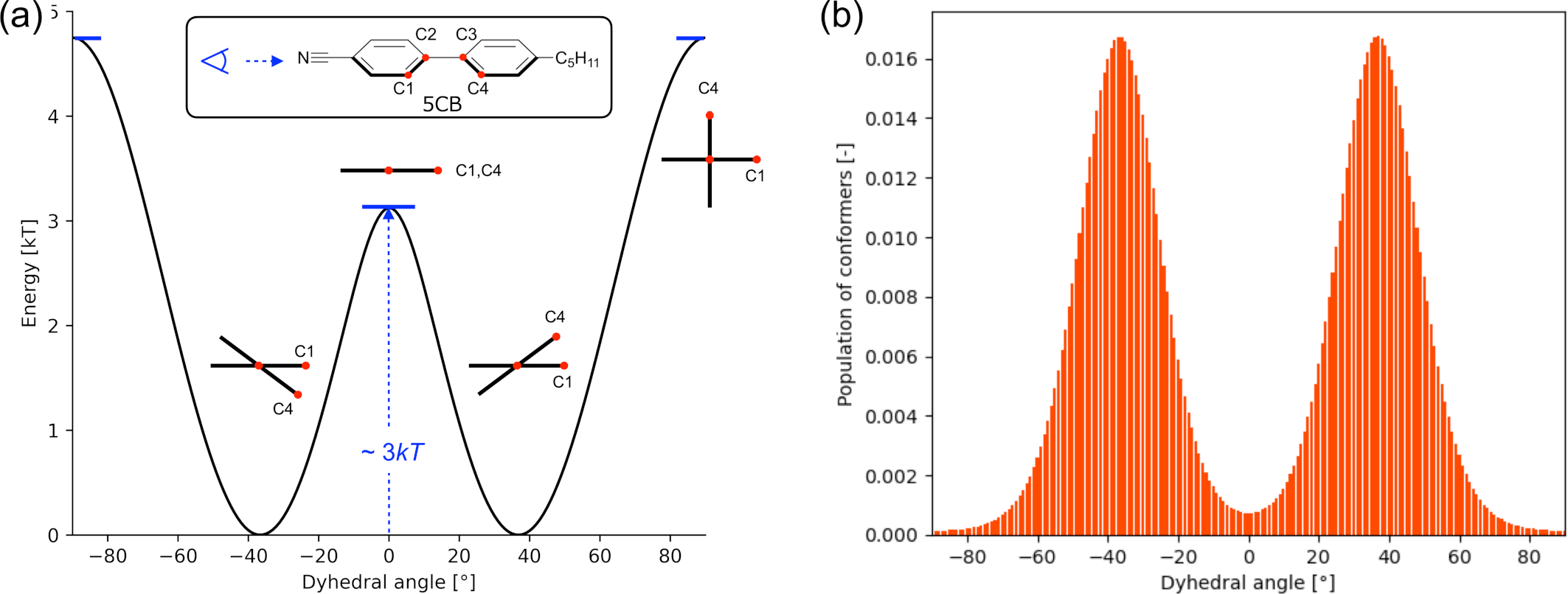}
  \caption{The chirality of the 5CB biphenyl group. (a) Strain energy on dihedral angles of the central bond of the biphenyl group of 5CB. (b) Boltzmann distribution of dihedral angles of the central bond of the biphenyl group of 5CB at 300K.}
  \label{fig:fig3}
\end{figure}

\subsection{Shape chirality of 5CB}
We performed the MD simulation of one of the most popular nematic LC compounds, 5CB, showing NLC phases at 300 K, with two types of chiral agents: ISO(6OBA)${}_2$ and 811 shown in Figure 3. ISO(6OBA)${}_2$ can be synthesized from a pharmaceutical isosorbide. Therefore, the other enantiomer is not comercially available. But, it is well-known for its strong helical twisting power (HTP). We simulated the mixtures of ISO(6OBA)${}_2$ and the other enantiomer. Meanwhile, we simulated the mixtures of the latter one because both the enantiomers, S811 and R811, are commercially available. They have weaker HTP than ISO(6OBA)${}_2$.  

MD simulations were performed on four mixtures, each of which contained approximately $1\%$ of both enantiomers of two types of chiral agents in a racemic mixture of 2048 molecules of 5CB. We calculated the dihedral angle at the C1-C2-C3-C4 shown in Figure 2a ($\phi$). The $\phi$ distribution of 5CB was examined for each mixture. For the two mixtures in which the chirality of the chiral agent was reversed, a difference in the average of $\phi$ ($\langle \phi \rangle$) emerged over time. As shown in Figure 4, it was found that the former increases the ratio of $P$-enantiomer of 5CB, and the latter increases the ratio of $M$-enantiomer of 5CB. The same trend was also observed for R811 and S811 with respect to chiral agent 811. In addition, the sign of the dihedral angles of 5CB was consistent with the sign of the circular polarization that causes the selective reflection exhibited by the real mixture, as listed in Table 1. As this is the situation after 200 ns at 300 K, it may be necessary to perform a long MD simulation. 

\begin{figure} 
  \centering
  \includegraphics[scale=0.5]{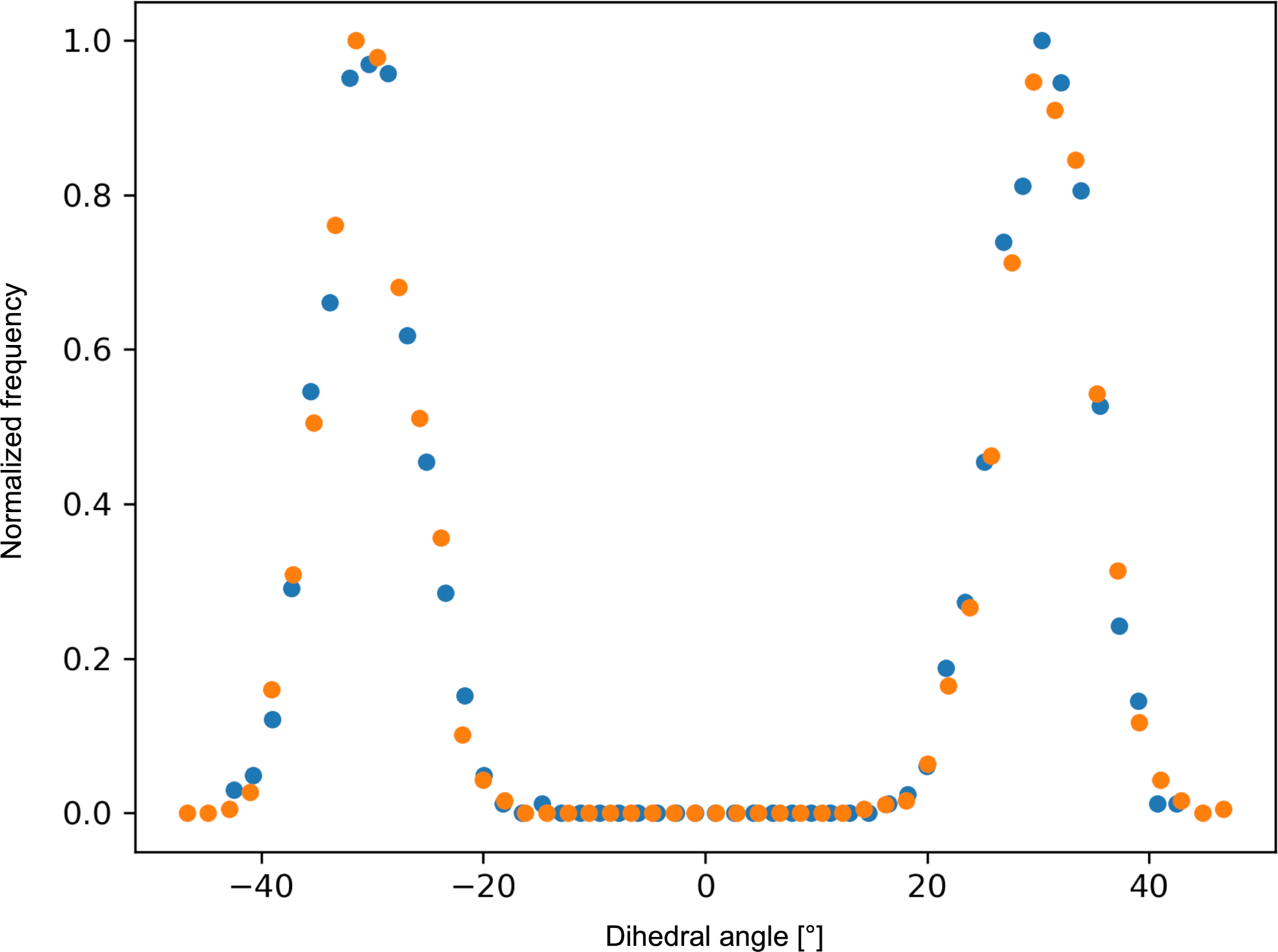}
  \caption{Histogram of dihedral angles of 2048 5CB molecules after 200 ns MD simulation with ISO(6OBA)${}_2$ (blue dots) and the other enantiomer (red dots).}
  \label{fig:fig4}
\end{figure}

\begin{table}
 \caption{Correlation between signs of selective reflection and molecular chirality}
  \centering
  \begin{tabular}{cccc}
    \toprule
    $Compound$ & Selective reflection & $\phi$ & $\chi_m$ \\
    \midrule
    R811 & R & $+$ & $-$\\
    S811 & L & $-$ & $+$\\
    ISO(6OBA)${}_2$ & R & $+$ & $+$\\
    The other enantiomer of ISO(6OBA)${}_2$ & L & $-$ & $-$\\
    \bottomrule
  \end{tabular}
  \label{tab:table1}
\end{table}

We tried to estimate the extent to which the transition between the two enantiomers of 5CB actually occurs and how long it takes to reach equilibrium. We calculated $\phi$ using the previous data of the all-atom MD simulation of 2009 molecules of 5CB with the same chirality. We calculated the time evolution of $\langle \phi \rangle$. When we took a $\phi$ histogram at 600 ns, we found that there were far more $P$-enantiomers, as shown in Figure 5a. Even after 600 ns, where the orientational order reached equilibrium, $\langle \phi \rangle$ had not yet reached 0, as shown in Figure 5a, which should be the equilibrium value. We calculated the time constant ($\tau$) of the equilibration using an exponential function, 

\begin{equation}
\langle \phi \rangle = \phi_0{\rm e}^{- t / \tau},
\end{equation}

where $\phi_0$ is $\langle \phi \rangle$ at the beginning of the equilibration. For the equilibration, $\phi_0$ and $\tau$ were estimated to be 30.5 and 9.04 $\mu \rm{s}$. Furthermore, since we started the MD situation in which all molecules are P enantiomers, the value of $\phi_0$ is reasonable. Meanwhile, $tau$ is unexpectedly long, considering that the system reached the equilibrium of the oriented structure at 600 ns. It can be further analyzed using the following equation.

\begin{equation}
\tau = \tau_0 {\rm e}^{E_0 / k_{\rm B} T}, 
\end{equation}

where $\tau_0$ means the limit when the temperature ($T$) is infinite, and $E_0$ is the energy barrier of the transition from $P$-enantiomer to $M$-enantiomer. $\tau_0$ is estimated to be 398 ${\rm ns}$ at 300 K using $E_0 = 3.12 k_{\rm B} T$, as shown in Figure 2a. Since $E$ is specific to the substance, $\tau_0$ is thought to represent the characteristic time for the transition between two states due to intermolecular interactions. It is not obvious whether there is a correlation between $\tau_0$ and the helical pitch when it reaches equilibrium, but it is predicted that the smaller $\tau_0$ is, the easier it is for the shape chirality shape, which is one of the intermolecular chirality information, to be transmitted. This suggests that the shape chirality shown in Figure 4 is not yet in equilibrium. At present, it is not cost-effective to perform MD simulations with a time scale of 10 $\mu \rm{s}$ or more. In the future, it will be necessary to work on efficient methods such as the replica exchange method. Here, however, we only report on the trends observed.

\begin{figure} 
  \centering
  \includegraphics[scale=0.8]{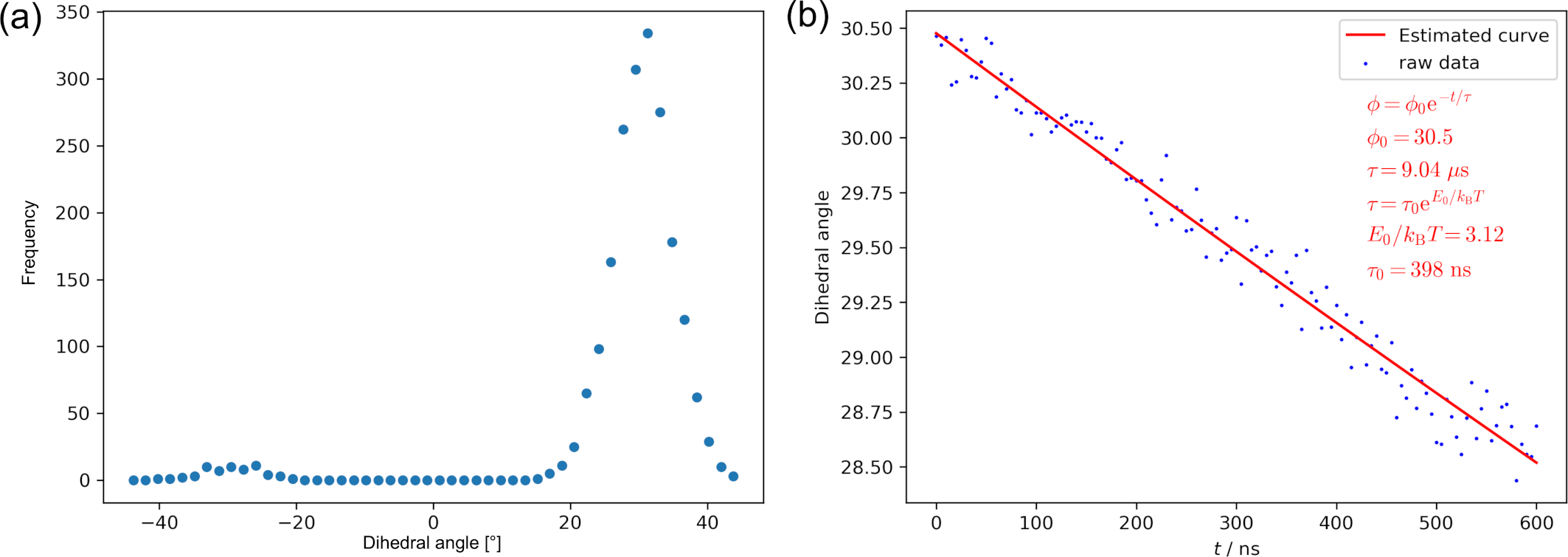}
  \caption{Equilibration process of $\langle \phi \rangle$ at 300K for 5CB. (a) Histogram of $\phi$ at 600 ns. (b) Time series of $\langle \phi \rangle$.}
  \label{fig:fig5}
\end{figure}

\subsection{Chirality of molecular motion of chiral agents}
Next, we would like to discuss the chirality of molecular motion. We have to define the chirality of molecular motion for each molecule. The motion obtained from MD simulations is the velocity of each atom at each time. In addition, molecules have mass, and we can consider them to be rigid bodies in snapshots. It is known that chirality can be defined for those for which momentum and angular momentum can be defined. Although various definitions with equivalent implications are possible, chirality can be defined as follows in this paper. First, the momentum vector $\vec{p}$ of an atom is the product of the velocity and atomic mass of the atom. The angular momentum vector $\vec{L}$ of an atom is the cross product of the position vector $\vec{r}$ of the atom in the coordinate system with the origin at the center of mass of the molecule and $\vec{p}$ of the same atom. The chirality of molecular motion, $\chi_{\rm m}$, is defined as the dot product of $\vec{p}$ and $\vec{L}$. The signs of $\chi_{\rm m}$ for the two enantiomers are opposite, as shown schematically in Figure 1c. Does simply mixing a chiral agent into an NLC phase with very large fluctuations cause the distribution of $\chi_{\rm m}$ to become biased in either direction?

We could not detect the chirality of molecular motion for 5CB. For the two pairs of ISO(6OBA)${}_2$ and 811, a single broad peak was obtained. The chirality of molecular motion is slightly but definitely biased, as shown in Figure 6. We performed the fitting of the histograms for both ISO(6OBA)${}_2$ and the other enantiomer to Voigt functions. For the mixture of ISO(6OBA)${}_2$ and the other enantiomer, the central $\chi_{\rm m}$ values were estimated to be $0.01379$ and $-0.01753$. The central $\chi_{\rm m}$ values were estimated to be $-0.237$ for R811 and $0.145$ for S811. 

\begin{figure} 
  \centering
  \includegraphics[scale=0.8]{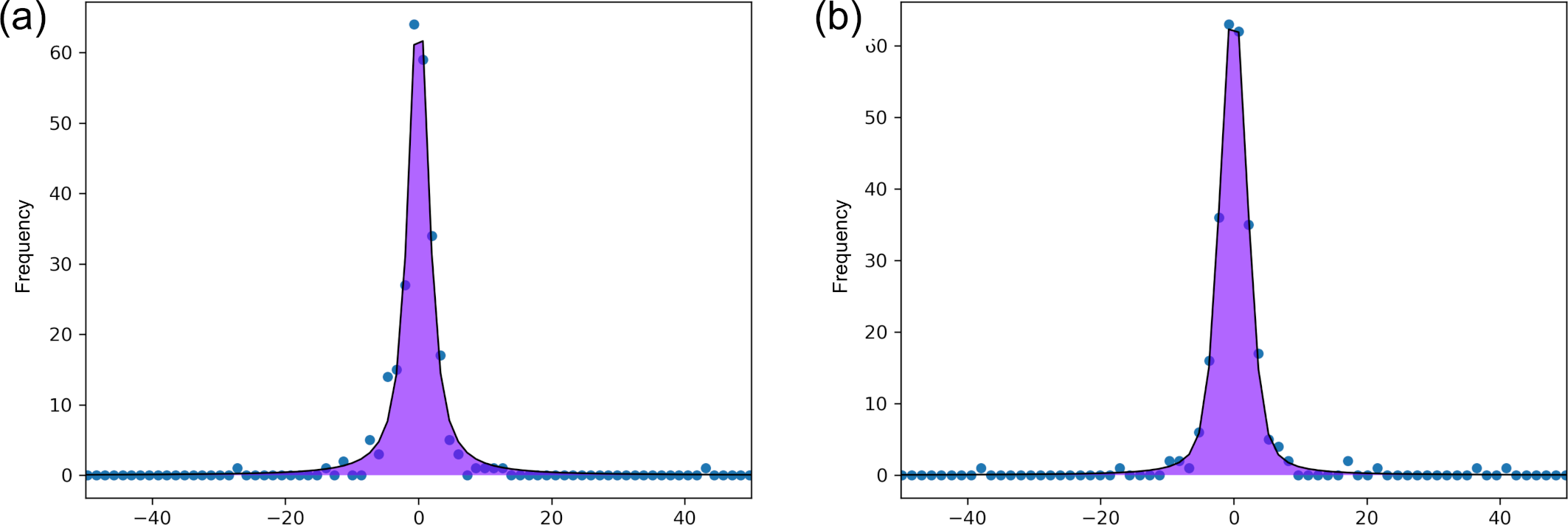}
  \caption{Histograms of $\chi_m$ of (a) ISO(6OBA)${}_2$ and (b) the other enantiomer for each MD simulation of 2048 5CB molecules with 1 wt\% chiral agent at 300 K.}
  \label{fig:fig6}
\end{figure}

The motion of the chiral agents is likely to affect the shape of 5CB. However, the chirality of the molecular motion of the chiral agents does not correlate with the shape chirality of the 5CB, as shown in Table 1. This means that the interaction of chiral agents affecting host NLC molecules is not uniquely determined by the sign of chiralities of molecular motion or shape alone. The reversal of helicity with temperature change is probably due to its complexity.

\section{Conclusion}
\label{sec:others}
It is well known that chiral molecules have shape chirality \cite{1998ferrarini_assessmentmolecularchirality}. This can be treated quantitatively using chirality parameters. The results obtained in this paper show that when a chiral agent is added to an achiral host NLC material, the shape chirality of the chiral agent is converted into the chirality of molecular motion in the host NLC molecules. We also found that this is transferred to the shape of the host NLC molecules, 5CB, as shown in Figure 7. This is the first confirmation of the origin of the interactions that induce intermolecular torsion as the mechanism for stabilizing CLC phases at the atomic level, which molecular theories have postulated \cite{1971gossens_MolecularTheoryCholestericPhaseTwistingPowerOpticallyActiveMoleculesNematicLiqudCrystal, 1976lin-liu_Molecularmodelcholestericliquidcrystals}. We thought that the motion of 5CB would also be transferred, but this could not be detected. Besides, we also found that it takes a very long time for the shape chirality of 5CB to reach equilibrium. Therefore, the chirality of the molecular motion of 5CB might become detectable when it is closer to equilibrium. Further investigation will be needed in this regard.

\begin{figure} 
  \centering
  \includegraphics[scale=0.5]{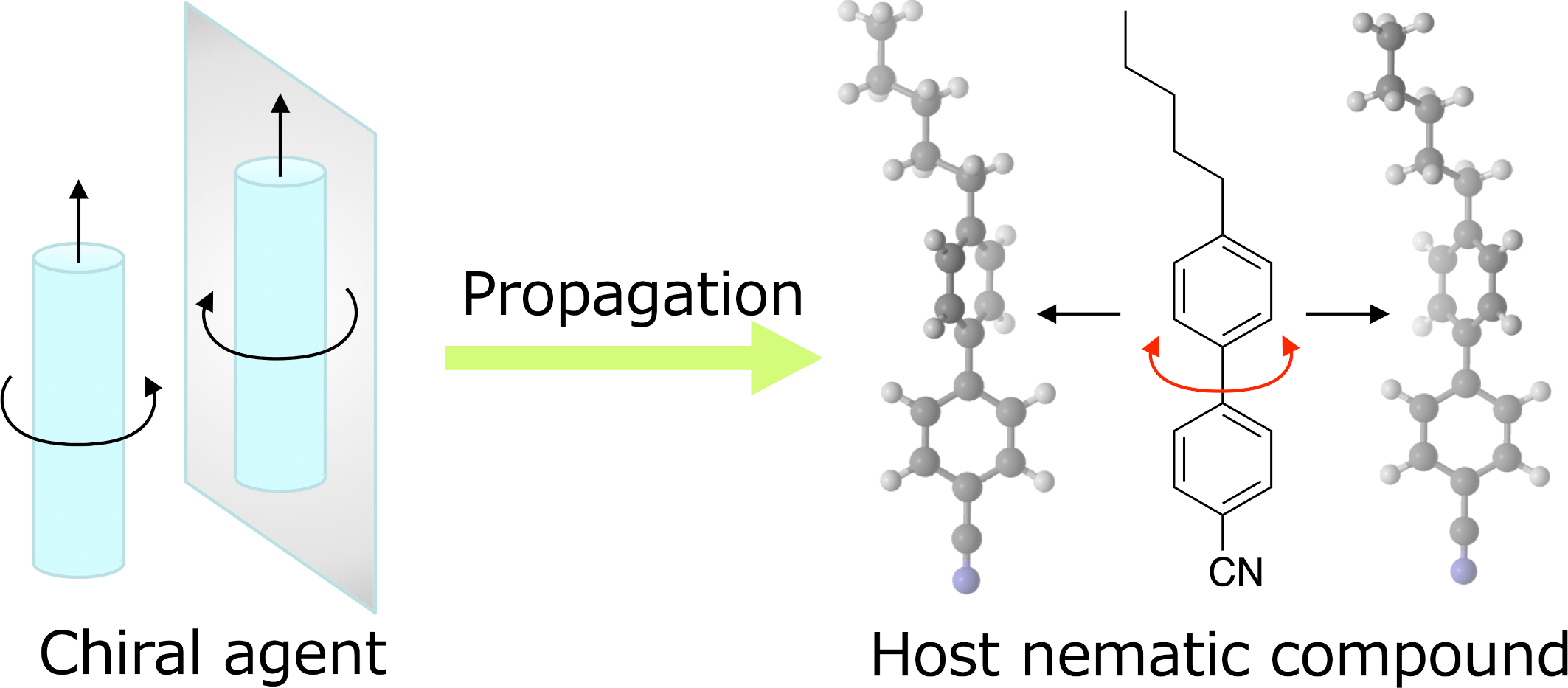}
  \caption{Mechanism of the chirality propagation from chiral agents to host nematic compounds. The shape chirality of chiral agents induces the chirality of molecular motion in a nemetic phase, and it induces the shape chirality of the host compounds.}
  \label{fig:fig7}
\end{figure}

\section{Acknowledgements}
\label{sec:others}
This work was supported in part by JSPS KAKENHI Grant Number JP23K23426. The computation was performed using Research Center for Computational Science, Okazaki, Japan (Project: 24-IMS-C050). The authors extend appreciation to Professor Yoshihiko Togawa, Osaka Metropolitan University, for helpful advice.

\section{Disclosure statement}
\label{sec:others}
No potential conflict of interest was reported by the authors.

\bibliographystyle{unsrt} 
\bibliography{Myself}  

\begin{thebibliography}{10}

\bibitem{2024lago-silva_Stimuliresponsivesynthetichelicalpolymers}
Mar{\'i}a {Lago-Silva}, Manuel {Fern{\'a}ndez-M{\'i}guez}, Rafael
  Rodr{\'i}guez, Emilio Qui{\~n}o{\'a}, and F{\'e}lix Freire.
\newblock Stimuli-responsive synthetic helical polymers.
\newblock {\em Chemical Society Reviews}, 53(2):793--852, 2024.

\bibitem{2024dreyfus_Dexteroushelicalmagneticrobotimprovedendovascularaccess}
R.~Dreyfus, Q.~Boehler, S.~Lyttle, P.~Gruber, J.~Lussi, C.~Chautems,
  S.~Gervasoni, J.~Berberat, D.~Seibold, N.~{Ochsenbein-K{\"o}lble},
  M.~Reinehr, M.~Weisskopf, L.~Remonda, and B.~J. Nelson.
\newblock Dexterous helical magnetic robot for improved endovascular access.
\newblock {\em Science Robotics}, 9(87):eadh0298, February 2024.

\bibitem{2023rey-tarrio_Multichiralmaterialscomprisingmetallosupramolecularcovalenthelicalpolymerscontainingfiveaxialmotifshelix}
Francisco {Rey- Tarr{\'i}o}, Emilio Qui{\~n}o{\'a}, Gustavo Fern{\'a}ndez, and
  F{\'e}lix Freire.
\newblock Multi-chiral materials comprising metallosupramolecular and covalent
  helical polymers containing five axial motifs within a helix.
\newblock {\em Nature Communications}, 14(1):3348, June 2023.

\bibitem{2023aizawa_Enantioselectivitydiscretizedhelicalsupramoleculeconsistingachiralcobaltphthalocyanineschiralinducedspinselectivityeffect}
Hiroki Aizawa, Takuro Sato, Saori {Maki-Yonekura}, Koji Yonekura, Kiyofumi
  Takaba, Tasuku Hamaguchi, Taketoshi Minato, and Hiroshi~M. Yamamoto.
\newblock Enantioselectivity of discretized helical supramolecule consisting of
  achiral cobalt phthalocyanines via chiral-induced spin selectivity effect.
\newblock {\em Nature Communications}, 14(1):4530, July 2023.

\bibitem{2023ren_SingleEnantiomerEmitterEnabledSuperstructuralHelixInversionUpconvertingDownshiftingLuminescenceBidirectionalCircularPolarization}
Chao Ren, Wenjing Sun, Tonghan Zhao, Chengxi Li, Chengyu Jiang, and Pengfei
  Duan.
\newblock A {{Single-Enantiomer Emitter Enabled Superstructural Helix
  Inversion}} for {{Upconverting}} and {{Downshifting Luminescence}} with
  {{Bidirectional Circular Polarization}}.
\newblock {\em Angewandte Chemie International Edition}, 62(50):e202315136,
  2023.

\bibitem{2002gennes_PhysicsLiquidCrystals}
P.~G. de~Gennes.
\newblock {\em {The Physics of Liquid Crystals}}.
\newblock Oxford University Press, U.S.A., Oxford, 2nd edition edition, May
  2002.

\bibitem{1998ferrarini_assessmentmolecularchirality}
Alberta Ferrarini and Pier~Luigi Nordio.
\newblock On the assessment of molecular chirality.
\newblock {\em Journal of the Chemical Society, Perkin Transactions 2},
  (2):455--460, January 1998.

\bibitem{2016watanabe_MolecularDynamicsApproachPredictingHelicalTwistingPowersMetalComplexDopantsNematicSolvents}
Go~Watanabe and Jun Yoshida.
\newblock Molecular {{Dynamics Approach}} for {{Predicting Helical Twisting
  Powers}} of {{Metal Complex Dopants}} in {{Nematic Solvents}}.
\newblock {\em The Journal of Physical Chemistry B}, 120(27):6858--6864, July
  2016.

\bibitem{2010anderson_sergeantsandsoldierseffectchiralamplificationnaphthalenediimidenanotubes}
Tom~W. Anderson, Jeremy K.~M. Sanders, and G.~Dan Panto{\c s}.
\newblock The sergeants-and-soldiers effect: Chiral amplification in
  naphthalenediimide nanotubes.
\newblock {\em Organic \& Biomolecular Chemistry}, 8(19):4274--4280, September
  2010.

\bibitem{2024zhou_Differentiatingenantiomersdirectionalrotationionsmassspectrometer}
Xiaoyu Zhou, Zhuofan Wang, Shuai Li, Xianle Rong, Jiexun Bu, Qiang Liu, and
  Zheng Ouyang.
\newblock Differentiating enantiomers by directional rotation of ions in a mass
  spectrometer.
\newblock {\em Science}, 383(6683):612--618, February 2024.

\bibitem{2023ishito_TrulychiralphononsaHgS}
Kyosuke Ishito, Huiling Mao, Yusuke Kousaka, Yoshihiko Togawa, Satoshi Iwasaki,
  Tiantian Zhang, Shuichi Murakami, Jun-ichiro Kishine, and Takuya Satoh.
\newblock Truly chiral phonons in {$\alpha$}-{{HgS}}.
\newblock {\em Nature Physics}, 19(1):35--39, January 2023.

\bibitem{2024oishi_Selectiveobservationenantiomericchiralphononsensuremathalphaquartz}
Eiichi Oishi, Yasuhiro Fujii, and Akitoshi Koreeda.
\newblock Selective observation of enantiomeric chiral phonons in
  \${\textbackslash}ensuremath\{{\textbackslash}alpha\}\$-quartz.
\newblock {\em Physical Review B}, 109(10):104306, March 2024.

\bibitem{2022kishine_DefinitionChiralityEnantioselectiveFields}
Jun-ichiro Kishine, Hiroaki Kusunose, and Hiroshi~M. Yamamoto.
\newblock On the {{Definition}} of {{Chirality}} and {{Enantioselective
  Fields}}.
\newblock {\em Israel Journal of Chemistry}, 62(11-12):e202200049, 2022.

\bibitem{2018nakagami_MolecularMobilityEffectMagneticInteractionsAllOrganicParamagneticLiquidCrystalNitroxideRadicalHydrogenBondingAcceptor}
Sho Nakagami, Takuya Akita, Daichi Kiyohara, Yoshiaki Uchida, Rui Tamura, and
  Norikazu Nishiyama.
\newblock Molecular {{Mobility Effect}} on {{Magnetic Interactions}} in
  {{All-Organic Paramagnetic Liquid Crystal}} with {{Nitroxide Radical}} as a
  {{Hydrogen-Bonding Acceptor}}.
\newblock {\em The Journal of Physical Chemistry B}, 122(29):7409--7415, July
  2018.

\bibitem{2020uchida_ThermalMolecularMotionCanAmplifyIntermolecularMagneticInteractions}
Yoshiaki Uchida, Go~Watanabe, Takuya Akita, and Norikazu Nishiyama.
\newblock Thermal {{Molecular Motion Can Amplify Intermolecular Magnetic
  Interactions}}.
\newblock {\em The Journal of Physical Chemistry B}, 124(28):6175--6180, July
  2020.

\bibitem{2019j.m.frisch_Gaussian16RevisionC02}
{J. M. Frisch}, {G. W. Trucks}, {H. B. Schlegel}, {G. E. Scuseria}, {M. A.
  Robb}, {J. R. Cheeseman}, {G. Scalmani}, {V. Barone}, {G. A. Petersson}, {H.
  Nakatsuji}, {X. Li}, {M. Caricato}, {A. V. Marenich}, {J. Bloino}, {B. G.
  Janesko}, {R. Gomperts}, {B. Mennucci}, {H. P. Hratchian}, {J. V. Ortiz}, {A.
  F. Izmaylov}, {J. L. Sonnenberg}, {D. Williams-Young}, {F. Ding}, {F.
  Lipparini}, {F. Egidi}, {J. Goings}, {B. Peng}, {A. Petrone}, {T. Henderson},
  {D. Ranasinghe}, {V. G. Zakrzewski}, {J. Gao}, {N. Rega}, {G. Zheng}, {W.
  Liang}, {M. Hada}, {M. Ehara}, {K. Toyota}, {R. Fukuda}, {J. Hasegawa}, {M.
  Ishida}, {T. Nakajima}, {Y. Honda}, {O. Kitao}, {H. Nakai}, {T. Vreven}, {K.
  Throssell}, {J. A. Montgomery, Jr.}, {J. E. Peralta}, {F. Ogliaro}, {M. J.
  Bearpark}, {J. J. Heyd}, {E. N. Brothers}, {K. N. Kudin}, {V. N. Staroverov},
  {T. A. Keith}, {R. Kobayashi}, {J. Normand}, {K. Raghavachari}, {A. P.
  Rendell}, {J. C. Burant}, {S. S. Iyengar}, {J. Tomasi}, {M. Cossi}, {J. M.
  Millam}, {M. Klene}, {C. Adamo}, {R. Cammi}, {J. W. Ochterski}, {R. L.
  Martin}, {K. Morokuma}, {O. Farkas}, {J. B. Foresman}, and {D. J. Fox}.
\newblock Gaussian 16, {{Revision C}}.02.
\newblock Gaussian, Inc., 2019.

\bibitem{1993bayly_wellbehavedelectrostaticpotentialbasedmethodusingchargerestraintsderivingatomicchargesRESPmodel}
Christopher~I. Bayly, Piotr Cieplak, Wendy Cornell, and Peter~A. Kollman.
\newblock A well-behaved electrostatic potential based method using charge
  restraints for deriving atomic charges: The {{RESP}} model.
\newblock {\em The Journal of Physical Chemistry}, 97(40):10269--10280, October
  1993.

\bibitem{2009m.j.frisch_Gaussian09RevisionD01}
{M. J. Frisch}, {G. W. Trucks}, {H. B. Schlegel}, {G. E. Scuseria}, {M. A.
  Robb}, {J. R. Cheeseman}, {G. Scalmani}, {V. Barone}, {B. Mennucci}, {G. A.
  Petersson}, {H. Nakatsuji}, {M. Caricato}, {X. Li}, {H. P. Hratchian}, {A. F.
  Izmaylov}, {J. Bloino}, {G. Zheng}, {J. L. Sonnenberg}, {M. Hada}, {M.
  Ehara}, {K. Toyota}, {R. Fukuda}, {J. Hasegawa}, {M. Ishida}, {T. Nakajima},
  {Y. Honda}, {O. Kitao}, {H. Nakai}, {T. Vreven}, {J. A. Montgomery}, {J. E.
  Peralta}, {F. Ogliaro}, {M. Bearpark}, {J. J. Heyd}, {E. Brothers}, {K. N.
  Kudin}, {V. N. Staroverov}, {R. Kobayashi}, {J. Normand}, {K. Raghavachari},
  {P. G. Rendell}, {J. C. Burant}, {S. S. Iyengar}, {J. Tomasi}, {M. Cossi},
  {N. Rega}, {J. M. Millam}, {M. Klene}, {J. E. Knox}, {J. B. Cross}, {V.
  Bakken}, {C. Adamo}, {J. Jaramillo}, {R. Gomperts}, {R. E. Stratmann}, {O.
  Yazyev}, {A. J. Austin}, {R. Cammi}, {C. Pomelli}, {J. W. Ochterski}, {R. L.
  Martin}, {K. Morokuma}, {V. G. Zakrzewski}, {G. A. Voth}, {P. Salvador}, {J.
  J. Dannenberg}, {S. Dapprich}, {A. D. Daniels}, {O. Farkas}, {J. B.
  Foresman}, {J. V. Ortiz}, {J. Cioslowski}, and {D. J. Fox}.
\newblock Gaussian 09 {{Revision D}}.01.
\newblock Gaussian, Inc., 2009.

\bibitem{2004wang_Developmenttestinggeneralamberforcefield}
Junmei Wang, Romain~M. Wolf, James~W. Caldwell, Peter~A. Kollman, and David~A.
  Case.
\newblock Development and testing of a general amber force field.
\newblock {\em Journal of Computational Chemistry}, 25(9):1157--1174, 2004.

\bibitem{1984berendsen_Moleculardynamicscouplingexternalbath}
H.~J.~C. Berendsen, J.~P.~M. Postma, W.~F. {van Gunsteren}, A.~DiNola, and
  J.~R. Haak.
\newblock Molecular dynamics with coupling to an external bath.
\newblock {\em The Journal of Chemical Physics}, 81(8):3684--3690, October
  1984.

\bibitem{1984nose_moleculardynamicsmethodsimulationscanonicalensemble}
Sh{\=u}ichi Nos{\'e}.
\newblock A molecular dynamics method for simulations in the canonical
  ensemble.
\newblock {\em Molecular Physics}, 52(2):255--268, June 1984.

\bibitem{1985hoover_CanonicaldynamicsEquilibriumphasespacedistributions}
William~G. Hoover.
\newblock Canonical dynamics: {{Equilibrium}} phase-space distributions.
\newblock {\em Physical Review A}, 31(3):1695--1697, March 1985.

\bibitem{1981parrinello_Polymorphictransitionssinglecrystalsnewmoleculardynamicsmethod}
M.~Parrinello and A.~Rahman.
\newblock Polymorphic transitions in single crystals: {{A}} new molecular
  dynamics method.
\newblock {\em Journal of Applied Physics}, 52(12):7182--7190, December 1981.

\bibitem{1997hess_LINCSlinearconstraintsolvermolecularsimulations}
Berk Hess, Henk Bekker, Herman J.~C. Berendsen, and Johannes G. E.~M. Fraaije.
\newblock {{LINCS}}: {{A}} linear constraint solver for molecular simulations.
\newblock {\em Journal of Computational Chemistry}, 18(12):1463--1472, 1997.

\bibitem{1971gossens_MolecularTheoryCholestericPhaseTwistingPowerOpticallyActiveMoleculesNematicLiqudCrystal}
W.~J.~A. Gossens.
\newblock A {{Molecular Theory}} of the {{Cholesteric Phase}} and of the
  {{Twisting Power}} of {{Optically Active Molecules}} in a {{Nematic Liqud
  Crystal}}.
\newblock {\em Molecular Crystals and Liquid Crystals}, 12(3):237--244,
  February 1971.

\bibitem{1976lin-liu_Molecularmodelcholestericliquidcrystals}
Y.~R. {Lin-Liu}.
\newblock Molecular model for cholesteric liquid crystals.
\newblock {\em Physical Review A}, 14(1):445--450, 1976.

\end{thebibliography}

\end{document}